\documentclass[a4paper,12pt]{article}

\usepackage{graphics, epsfig}

\begin{document}

\author{C. Barrab\`es\thanks{E-mail : barrabes@celfi.phys.univ-tours.fr}\\     
\small Laboratoire de Math\'ematiques et Physique Th\'eorique\\
\small  CNRS/UPRES-A 6083, Universit\'e F. Rabelais, 37200 TOURS, France\\
P.A. Hogan\thanks{E-mail : phogan@ollamh.ucd.ie}\\
\small Mathematical Physics Department\\
\small  National University of Ireland Dublin, Belfield, Dublin 4, Ireland}

\title{Recoil and Ring--Down Effects in Gravitation}
\date{}
\maketitle

\begin{abstract}
We construct a model of a relativistic fireball or light--like 
shell of matter by considering a spherically symmetric moving 
gravitating mass experiencing an impulsive deceleration to rest. 
We take this event to be followed by the mass undergoing a 
deformation leading to the emission of gravitational radiation while 
it returns to a spherically symmetric state. We find that the 
fireball is accompanied by an impulsive gravitational wave.

\end{abstract}
\thispagestyle{empty}
\newpage

\section{Introduction}\indent
In a recent paper \cite{1} we constructed a model in general 
relativity of a spherically symmetric, moving gravitating mass 
which experiences an impulsive deceleration to rest, as viewed 
by a distant observer. To the distant observer the mass, which 
is moving rectilinearly with uniform 3--velocity $v$, suddenly 
halts. This event is accompanied by the emergence of a spherical 
light--like shell whose total energy measured by the observer is 
found to be the same as the relative kinetic energy of the source 
before it stops. The shell may be considered a `relativistic 
fireball' and such objects are believed to constitute the 
`inner engine' of the fireball model of gamma--ray bursts \cite{2}. 
The deceleration phenomenon is a recoil effect and may be thought 
of as a limiting case of a Kinnersley rocket \cite{3} ,\cite {4}. 
It has an electromagnetic analogue in which the mass is replaced 
by a point charge and the light--like shell is replaced by a 
spherical impulsive electromagnetic wave \cite{1}.

It was pointed out in \cite{1} that the concept of an extended spherical 
body remaining rotationally symmetric after suddenly decelerating to 
rest is a strong idealisation. It would be more realistic to expect the 
body to experience some deformation following the deceleration which 
would lead to the emission of gravitational radiation before eventually 
settling back to a spherically symmetric state. This is the question 
we address in the present paper. To do so we make use of the Schwarzschild 
space--time, with mass parameter $m$ and with parameter $v$ representing 
the uniform 3--velocity of the spherical source viewed by a distant 
observer, and the Robinson space--time \cite{5} describing the vacuum 
gravitational field of an axially symmetric, isolated body 
which rapidly evolves into a Schwarzschild space--time due to the 
emission of gravitational radiation. Both of these space--times contain 
future null--cones (i.e. null hypersurfaces generated by shear--free, 
expanding 
null geodesics) and we attach them to each other in an appropriate manner 
on one such null--cone ${\cal N}$, 
with the two--parameter Schwarzschild space--
time to the past of ${\cal N}$ and the Robinson space--time to the future of 
${\cal N}$. We thus create the more realistic model of sudden deceleration to 
rest which we set out to do and the remaining question is: what is the 
physical nature of the light--like signal whose history in space--time 
is the future null--cone ${\cal N}$? 
We show that {\it ${\cal N}$ is the history of 
a light--like shell accompanied by an impulsive gravitational wave}. The 
total energy of the shell measured by the distant observer refered to 
above is calculated as a perturbation of its value obtained in \cite{1} 
when the deceleration was instantly followed by a Schwarzschild space--
time with $v=0$. If this energy is {\it less} than the original kinetic energy 
of the source then the sign of a small parameter involved in the 
approximate calculation is determined and this in turn means that the 
signal front is a prolate spheroid which, in the ring--down phase, 
quickly becomes a sphere and the static Schwarzschild space--time is 
then established.

The paper is organised as follows: in section 2 the gravitational 
radiation--free model of sudden deceleration to rest of a Schwarzschild 
source having uniform 3--velocity $v$ is described. This is followed 
in section 3 by the more realistic model in which, after deceleration 
to rest, the source is an axially symmetric radiating source whose 
gravitational field is described by the Robinson space--time. The paper 
ends with a discussion in which the non--spherical shape of the wave 
front in the ring--down phase is examined.

\setcounter{equation}{0}
\section{The Gravitational Radiation--Free Model}\indent
To set the scene we review the spherically symmetric model 
of a recoil effect \cite{1}. Consider first the Schwarzschild 
line--element in the form 
\begin{equation}\label{2.1}
ds^2=k^2r^2\left\{\frac{d\xi ^2}{1-\xi ^2}+(1-\xi ^2)\,d\phi ^2\right\}
-2du\,dr-\left (1-\frac{2\,m}{r}\right )\,du^2\ ,
\end{equation}
with $k^{-1}=\gamma\,(1-v\,\xi )\ , \gamma =(1-v^2)^{-1/2}$ and 
$0\leq v <1$. Here the coordinate ranges are $-1\leq \xi\leq +1\ , 
0\leq\phi <2\,\pi\ , -\infty <u<+\infty$ and $0\leq r<+\infty$. Instead 
of using the polar angle $\theta$ we have used $\xi =\cos\theta$ as a 
coordinate. This proves to be useful for all subsequent calculations. 
The constant $m$ is the mass of the source and the constant $v$ is the 3--velocity 
of the source which appears to be moving rectilinearly to a distant 
observer. The line--element (\ref{2.1}) can be transformed to the 
usual 1--parameter Schwarzschild form
\begin{equation}\label{2.2}
ds^2={\bar r^2}\left\{\frac{d{\bar \xi}^2}{1-{\bar \xi}^2}+(1-{\bar \xi}^2)
\,d{\bar \phi}^2\right\}
-2d{\bar u}\,d{\bar r}-\left (1-\frac{2\,m}{{\bar r}}\right )\,d{\bar u}^2\ ,
\end{equation}
by the transformation \cite{1}
\begin{equation}\label{2.3}
{\bar \xi}=\frac{\xi -v}{1-v\,\xi}\ ,\qquad {\bar \phi}=\phi\ ,\qquad 
{\bar r}=r\ ,\qquad {\bar u}=u\ .
\end{equation}
From the form (\ref{2.2}) we conclude that $m$ is the Bondi mass or 
`mass aspect' in this case of the source at rest \cite{6}.
By writing (\ref{2.1}) in Bondi form we will find 
the mass aspect associated with a mass $m$ 
moving along the symmetry axis with 3--velocity $v$ [cf. \cite{6}, equation (72)]. 
This confirms the physical interpretation of the parameter $v$. To see this 
we begin with the exact transformation
\begin{eqnarray}\label{2.4}
r' & = & k\,r\,\varphi\ ,\\
u' & = & \gamma\,u+k\,r\,(1-\varphi )\ ,\\
\xi ' & = & \varphi ^{-1}\left (\chi +\xi\right )\ ,\\
\phi ' & = & \phi\ ,
\end{eqnarray}
with
\begin{equation}\label{2.8}
\varphi =\left (1+2\,\chi\,\xi +\chi ^2\right )^{1/2}\ ,\qquad 
\chi =\frac{\gamma\,v\,u\,}{k\,r}\ ,
\end{equation}
The transformation (2.4)--(2.8) appeared first in \cite{1} and 
is utilised there for a different purpose. Its effect on (\ref{2.1}) 
is to yield
\begin{equation}\label{2.9}
ds^2=r'^2\left\{\frac{d\xi '^2}{1-\xi '^2}+(1-\xi '^2)\,d\phi '^2\right\}
-2du'\,dr'-du'^2+\frac{2\,m\,k\,\varphi}{r'}\,du^2\ .
\end{equation}
We will simplify the last term here assuming $r'$ is large. To this end 
we first find that for large $r'$, 
\begin{equation}\label{2.10}
\varphi =1+\frac{\gamma\,v\,u\,\xi '}{r'}+O\left (\frac{1}{r'^2}\right )\ ,
\end{equation}
and so we obtain
\begin{eqnarray}\label{2.11}
\xi & = & \xi '-\frac{\gamma\,v\,k'\,u'\,(1-\xi '^2)}{r'}+O\left (\frac{1}{r'^2}\right )\ ,\\
\phi & = & \phi '\ ,\\
r & = & \frac{r'}{k'}+\gamma ^2v\,k'\,u'\,(v-\xi ')+O\left (\frac{1}{r'}\right )\ ,\\
u & = & k'\,u'+O\left (\frac{1}{r'}\right )\ ,
\end{eqnarray}
with $k'^{-1}=\gamma\,(1-v\,\xi ')$. Now (2.9) becomes
\begin{eqnarray}\label{2.15}
ds^2 & = & r'^2\left\{\frac{d\xi '^2}{1-\xi '^2}+(1-\xi '^2)\,d\phi '^2\right\}-2du'\,dr'-du'^2 \nonumber\\
{} & + & \frac{2\,m\,k'^3}{r'}\,(du'+k'\,u'\,\gamma\,v\,d\xi '
)^2+O\left (\frac{1}{r'^2}\right )\ .
\end{eqnarray}
A further transformation $r'\rightarrow r''$ given by
\begin{equation}\label{2.17}
r''=r'+\frac{m\,k'^5\gamma ^2\,v^2u'^2(1-\xi '^2)}{2\,r'^2}+
O\left (\frac{1}{r'^3}\right )\ ,
\end{equation}
puts the line--element (2.15) in Bondi form \cite{6},
\begin{eqnarray}\label{2.18}
ds^2 & = & r''^2\left\{\frac{{\rm e}^{2\lambda}}{(1-\xi '^2)}
\left (d\xi '+U\,\sqrt{1-\xi '^2}\,du'\right )^2+{\rm e}^{-2\lambda}
\,(1-\xi '^2)\,d\phi '^2\right\}\nonumber\\
&  &-2\,{\rm e}^{2\beta}du'\,dr''
-r''^{-1}V\,{\rm e}^{2\beta}du'^2\ ,
\end{eqnarray}
with
\begin{eqnarray}\label{2.19}
\lambda & = & \frac{M\,\gamma ^2\,v^2\,u'^2\,k'^2\,
(1-\xi '^2)}{2\,r''^3}+O\left (\frac{1}{r''^4}\right )\ ,\\
U & = & \frac{2\,M\,\gamma\,v\,u'\,k'\,\sqrt{1-\xi '^2}}{r''^3}+O\left (\frac{1}{r''^4}\right )\ ,\\
\beta & = & O\left (\frac{1}{r''^2}\right )\ ,\\
V & = & r''-2\,M+O\left (\frac{1}{r''}\right )\ ,
\end{eqnarray}
and
\begin{equation}\label{2.23}
M=m\,k'^3\ .
\end{equation}
Thus $M$ is the `mass aspect' of the moving source. This is precisely the mass 
aspect associated with a mass $m$ moving with velocity $v$ along the axis 
of symmetry \cite{6}.

To set up our model describing the instantaneous deceleration to rest of a Schwarzschild 
source we consider the Schwarzschild space--time ${\cal M}$ subdivided into 
two halves $\cal M^+$ and $\cal M^-$ each with boundary the future null--cone 
${\cal N}\,(u=0)$. For the line--element of the region $\cal M^-$, which we take to be the 
past of ${\cal N}$, we use (2.1). For the line--element of the region $\cal M^+$ to 
the future of ${\cal N}$ we take 
\begin{equation}\label{2.24}
ds_+^2=r_+^2\left\{\frac{d\xi _+^2}{1-\xi _+^2}+(1-\xi _+^2)\,d\phi ^2\right\}
-2du\,dr_+-\left (1-\frac{2\,m}{r_+}\right )\,du^2\ .
\end{equation}
These space--times are attached on ${\cal N}$ with the matching conditions
\begin{equation}\label{2.25}
\xi _+=\xi\ ,\qquad r_+=k\,r\ ,
\end{equation}
with $k^{-1}=\gamma\,(1-v\,\xi )$, which will ensure that the induced line--element 
on ${\cal N}$ from its embedding in $\cal M^+$ 
agrees with the induced line--element 
on ${\cal N}$ from its embedding in $\cal M^-$. In this way we have constructed the 
space--time denoted $\cal M^-\cup\cal M^+$. Other ways of mapping 
${\cal N}$ to itself 
preserving the induced line--element are possible and lead to different 
models. The particular matching (2.25) is motivated by the electromagnetic 
analogue of this gravitational deceleration problem in \cite{1}.

We now use the theory of light--like signals in general relativity developed 
by Barrab\`es and Israel (denoted BI) \cite{7} to discover the physical 
properties of the signal with history ${\cal N}$. In general the Einstein tensor 
and the Weyl tensor of the space--time $\cal M^-\cup\cal M^+$ will each 
have a Dirac $\delta (u)$--term and the BI theory enables us to calculate 
its coefficient using (2.1), (2.24) and the matching conditions (2.25). If 
the coefficient of $\delta (u)$ in the Einstein tensor is non--zero then it 
is simply associated with a surface stress--energy for the light--like shell. 
If the coefficient of $\delta (u)$ in the Weyl tensor has a part which is 
type N in the Petrov classification with degenerate principal null direction 
coinciding with the direction of the null normal to $u=0$ then the signal 
with history ${\cal N}\,(u=0)$ contains an impulsive gravitational wave. This 
is discussed in \cite{7}, \cite {8}. For the reader familiar with the 
BI theory we now provide a guide through the present application.    

The local coordinate system in ${\cal M}^-$ with line--element (\ref{2.1}) 
is denoted $\{x^\mu _-\}=\{\xi , \phi , r, u\}$ while the local 
coordinate system in ${\cal M}^+$ with line--element (\ref{2.24}) is 
denoted $\{x^\mu _+\}=\{\xi _+, \phi , r_+, u\}$. As normal 
to ${\cal N}$ we take the null vector field with components $n_\mu$ given via 
the 1--form $n_\mu dx^\mu _{\pm}=-du$. Since we want the physical 
properties of ${\cal N}$ observed by the observer using the 
plus coordinates we take as intrinsic coordinates on 
${\cal N}$, $\{\xi ^a\}=\{\xi _+, \phi , r_+\}$ with $a=1, 2, 3$. 
A set of three linearly independent tangent vector fields to ${\cal N}$ 
is $\left\{e_{(1)}=
\partial /\partial\xi _+, e_{(2)}=\partial /\partial\phi , 
e_{(3)}=\partial /\partial r_+\right\}$. The components of these 
vectors on the plus side of ${\cal N}$ are 
$e^\mu _{(a)}|_+=\delta ^\mu _a$. The components of these 
vectors on the minus side of ${\cal N}$ are 
\begin{equation} \label{2.26}
e^\mu _{(a)}|_-=\frac{\partial x^\mu _-}{\partial\xi ^a}\ ,
\end{equation}
with the relation between $\{x^\mu _-\}$ and $\{\xi ^a\}$ 
given by the {\it matching conditions} (\ref{2.25}). Hence 
we find that
\begin{eqnarray} \label{2.27}
e^\mu _{(1)}|_- &=& (1, 0,-r_+\gamma\,v, 0)\ ,\\
e^\mu _{(2)}|_- &=& (0, 1, 0, 0)\ ,\\
e^\mu _{(3)}|_- &=& (0, 0, \gamma\,(1-v\,\xi _+), 0)\ .
\end{eqnarray}
We need a transversal on ${\cal N}$ consisting of a vector field on 
${\cal N}$ which points out of ${\cal N}$. A convenient such (covariant) vector 
expressed in the coordinates $\{x^\mu _+\}$ is ${}^+N_\mu =
(0, 0, 1, \frac{1}{2}-\frac{m}{r_+})$. Thus since $n^\mu =\delta ^\mu _3$ 
we have ${}^+N_\mu n^\mu =1$.We next construct the transversal on 
the minus side 
of ${\cal N}$ with covariant components ${}^-N_\mu$. To ensure that 
this is the same vector on the minus side of ${\cal N}$ as ${}^+N_\mu$ 
when viewed on the plus side we require
\begin{equation} \label{2.30}
{}^+N_ \mu\,e^\mu _{(a)}|_+={}^-N_ \mu\,e^\mu _{(a)}|_-\ ,
\qquad {}^+N_\mu {}^+N^\mu ={}^-N_\mu {}^-N^\mu\ .
\end{equation}
The latter scalar product is zero as we have chosen to 
use a null transversal. We find that
\begin{equation} \label{2.31}
{}^-N_\mu =\left (\frac{r_+v}{1-v\,\xi _+}, 0, \frac{1}
{\gamma\,(1-v\,\xi _+)}, D\right )\ ,
\end{equation}
with
\begin{equation} \label{2.32}
D=\frac{v^2(1-\xi _+^2)\,\gamma}{2(1-v\,\xi _+)}+
\frac{1}{2\gamma\,(1-v\,\xi _+)}-\frac{m}
{\gamma ^2(1-v\,\xi _+)^2r_+}\ .
\end{equation}
Next the transverse extrinsic curvature on the plus and minus 
sides of ${\cal N}$ is given by
\begin{equation} \label{2.33}
{}^{\pm}{\cal K}_{ab}=-{}^{\pm}N_\mu\left (\frac{\partial e^\mu 
_{(a)}|_{\pm}}{\partial\xi ^b}+{}^{\pm}\Gamma ^\mu _{\alpha\beta}\,
e^\alpha _{(a)}|_{\pm}e^\beta _{(b)}|_{\pm}\right )\ ,
\end{equation}
where ${}^{\pm}\Gamma ^\mu _{\alpha\beta}$ are the components 
of the Riemannian connection associated with the metric tensor of 
$\cal M^+$ or $\cal M^-$ evaluated on ${\cal N}$. The key quantity we need is 
the jump in the transverse extrinsic curvature across ${\cal N}$ given 
by
\begin{equation} \label{2.34}
\sigma _{ab}=2\,\left ({}^+{\cal K}_{ab}-{}^-{\cal K}_{ab}\right )
\ .
\end{equation}
This jump is independent of the choice of transversal on ${\cal N}$ 
\cite{7}. We find that in the present application $\sigma _{ab}=0$ 
except for 
\begin{equation} \label{2.35}
\sigma _{11}=\frac{2}{1-\xi _+^2}\,\left (m\,k^3-m\right )\ ,
\qquad \sigma _{22}=2\,(1-\xi _+^2)\left (m\,k^3-m\right )\ ,
\end{equation}
with $k^{-1} = \gamma (1-v \xi)$. 
Now $\sigma _{ab}$ is extended to a 4--tensor field on ${\cal N}$ with 
components $\sigma _{\mu\nu}$ by padding--out with zeros (the 
only requirement on $\sigma _{\mu\nu}$ is $\sigma _{\mu\nu}\,
e^\mu _{(a)}|_{\pm}\,e^\nu _{(b)}|_{\pm}=\sigma _{ab}$). With our 
choice of future--pointing normal to ${\cal N}$ and past--pointing 
transversal, the surface stress--energy tensor components are 
$-S_{\mu\nu}$ with $S_{\mu\nu}$ given by 
\cite{7}
\begin{equation} \label{2.36}
16\pi\,S_{\mu\nu}=2\,\sigma _{(\mu}\,n_{\nu )}-\sigma\,n_\mu\,n_\nu 
-\sigma ^{\dagger}g_{\mu\nu}\ ,
\end{equation}
with
\begin{equation} \label{2.37}
\sigma _\mu =\sigma _{\mu\nu}\,n^\nu\ ,\qquad \sigma ^{\dagger}
=\sigma _\mu\,n^\mu\ ,\qquad \sigma =g^{\mu\nu}\gamma _{\mu\nu}\ .
\end{equation}
In the present case $\sigma _\mu =0$ and thus $\sigma ^{\dagger}=0$ and 
the surface stress--energy tensor takes the form
\begin{equation} \label{2.38}
-S_{\mu\nu}=\rho\,n_\mu\,n_\nu\ .
\end{equation}
Hence the energy density of the light--like shell measured 
by the distant observer using the plus coordinates is \cite{7} 
\begin{equation} \label{2.39}
\rho =\frac{\sigma}{16\pi}=\frac{1}{4\pi\,r_+^2}\,
\left (m\,k^3-m\right )\ .
\end{equation}
Thus the null--cone ${\cal N}$ is the history of a light--like shell 
with surface stress--energy given by (\ref{2.38}). We 
note that $m\,k^3$ is the ``mass aspect" (cf. (\ref{2.23})) 
on the minus side of ${\cal N}$. 
A calculation of the singular $\delta$--part of the Weyl tensor 
for $\cal M^-\cup\cal M^+$ reveals that it vanishes. Hence there is 
no possibility of the light--like signal with history ${\cal N}$ containing 
an impulsive gravitational wave. We note that $\rho$ is 
a monotonically increasing function of $\xi _+$. Thus on the 
interval $-1\leq\xi _+\leq +1$, $\rho$ is maximum at $\xi _+=+1$ (in the direction of the 
motion) and $\rho$ is minimum at $\xi _+=-1$. This 
is as one would expect. A burst of null matter predominantly 
in the direction of motion is required to halt the mass. 
In this sense the model we 
have constructed here could be thought of as a limiting case of 
a Kinnersley rocket \cite{3} \cite{4}.

By integrating (\ref{2.39}) over the shell with 
area element $dA_+=r_+^2\,d\xi _+\,d\phi$ and with $-1\leq\xi _+\leq +1, 
0\leq\phi <2\pi$ we obtain the total energy $E_+$ of the shell 
measured by the distant observer who sees the mass $m$, moving 
rectilinearly with 3--velocity  $v$ in the direction $\xi _+=+1$, 
suddenly halted. Thus
\begin{equation} \label{2.40}
E_+=\frac{1}{4\pi}\,\int_{0}^{2\pi}d\phi _+\,\int_{-1}^{+1}
\left (m\,k^3-m\right )\,d\xi _+\ .
\end{equation}
This results in
\begin{equation} \label{2.41}
E_+=m\,(\gamma -1)\ .
\end{equation}
Thus all of the kinetic energy of the mass $m$ 
before stopping is converted into the relativistic shell. This 
is a satisfactory result from the point of view of energy conservation. 
It is interesting that an exact formula from special relativity has 
emerged from an exact calculation in general relativity.

\setcounter{equation}{0}
\section{Recoil with Ring--Down}\indent
In order to achieve a more realistic model we take the space--time 
$\cal M^+$ to be a Robinson space--time rather than the Schwarzschild 
space--time with line--element (\ref{2.24}). Thus (\ref{2.24}) is 
replaced by the line--element \cite{5}
\begin{equation}\label{3.1}
ds_+^2=r_+^2\{f_+^{-1}(d\xi _++a_+f_+du)^2+f_+d\phi ^2\}-2\,du\,dr_+
-c_+du^2\ ,
\end{equation}
where $f_+, a_+$ are functions of $\xi _+, u$ and 
\begin{equation}\label{3.2}
c_+=K_+-2\,H_+r_+-\frac{2\,m}{r_+}\ ,
\end{equation}
with $m$ a constant,
\begin{equation}\label{3.3}
H_+=\frac{1}{2}\,\left (a_+f_+\right )'\ ,\qquad K_+=-\frac{1}{2}\,f_+''\ ,
\end{equation}
and 
\begin{equation}\label{3.4}
a_+'=-f_+^{-2}\dot f_+\ ,\qquad \left (6\,m\,a_+f_++f_+K_+'\right )'
=0\ ,
\end{equation}
with the prime denoting differentiation with respect to $\xi _+$ 
and the dot denoting differentiation with respect to $u$. Here the 
coordinates $\{\xi _+, \phi , r_+, u\}$ have the ranges $-1\leq\xi _+
\leq +1,\  0\leq\phi <2\pi ,\  0<r_+<+\infty ,\  0\leq u<+\infty$ in $\cal M^+$. 
The hypersurfaces $u={\rm constant}$ are future null--cones (in 
the sense used above) and $\cal M^+$ is attached to $\cal M^-$, with 
line--element (\ref{2.1}), on the null--cone ${\cal N}(u=0)$ with the 
natural generalisation of the matching (\ref{2.25}) given by
\begin{equation}\label{3.5}
r_+=\frac{1}{g}\,r\ ,\qquad \frac{d\xi _+}{d\xi}=\frac{f_+}{1-\xi ^2}\ 
,\end{equation}
with
\begin{equation}\label{3.6}
g=k^{-1}\,\left (\frac{f_+}{(1-\xi ^2)}\right )^{1/2}\ ,
\end{equation}
evaluated on $u=0$ with $k^{-1}=\gamma\,(1-v\,\xi )$ as in (\ref{2.1}). 
The function $f_+(\xi _+, u)$ satisfies the conditions \cite{5}
\begin{equation}\label{3.7}
f_+(\pm 1, u)=0\ ,\qquad f_+'(-1, u)=2=-f_+'(+1, u)\ .
\end{equation}
The space--time with line--element (\ref{3.1}) and with (\ref{3.2})--
(\ref{3.4}) and (\ref{3.7}) holding is a vacuum space--time 
containing gravitational waves from a bounded source having smooth 
wave fronts which are homeomorphic to a 2--sphere. Robinson \cite{5} has 
shown very elegantly how such a space--time will evolve with increasing 
$u$ into a Schwarzschild space--time with mass $m$. He has in addition 
given an exact solution in the form of a power series for the function 
$f_+$ in powers of $\exp\left (-\frac{2\,u}{m}\right )$, with 
coefficients which are polynomials in $\xi _+$. The derived functions $a_+, 
K_+, H_+$ are then also given as similar power series in this variable. 
The function $f_+$ has the general form \cite{5}
\begin{equation}\label{3.8}
f_+=(1-\xi ^2)\,\left\{1+(1-\xi ^2)\,\hat f(\xi _+, u)\right\}\ ,
\end{equation}
with
\begin{equation}\label{3.9}
\hat f(\xi _+, u)=\sum_{n=0}^{\infty}c_n\,\hat f_n(\xi _+)\,
{\rm e}^{-2(n+1)\,u/m}\ ,
\end{equation}
where $c_n$ are arbitrary constants for $n=0, 1, 2, \dots$ and 
$\hat f_n(\xi _+)$ are known polynomials. We shall assume that when $u=0$ the 
constants $c_n$ are sufficiently small to ensure that the series of 
polynomials $\hat f_+(\xi _+, 0)$ converges uniformly for $-1\leq\xi _+
\leq +1$. The leading terms in the power series for $f_+, a_+, K_+, 
H_+$ for the Robinson solution are 
\begin{eqnarray}\label{3.10}
f_+ & = & 1-\xi _+^2+c_0(1-\xi _+^2)^2{\rm e}^{-2\,u/m}+\dots \ ,\\
a_+ & = & \frac{2\,c_0\xi _+}{m}\,{\rm e}^{-2\,u/m}+\dots \ ,\\
K_+ & = & 1-2\,c_0(3\,\xi _+^2-1)\,{\rm e}^{-2\,u/m}+\dots \ ,\\
H_+ & = & \frac{c_0}{m}\,(1-3\,\xi ^2_+)\,{\rm e}^{-2\,u/m}+\dots \ .
\end{eqnarray}
The calculation using the BI theory to establish the physical nature 
of the signal having as history the null-cone ${\cal N}(u=0)$ parallels that 
given in section 2. We will outline the differences between the results 
given in section 2 and the calculations in the present case. Firstly (2.26) 
remains unchanged while (2.27)--(2.29) are now 
replaced by
\begin{eqnarray}\label{3.14}
e^{\mu}_{(1)}|_- & = & (k^{-2}g^{-2}, 0, r_+g', 0)\ ,\\
e^{\mu}_{(2)}|_- & = & (0, 1, 0, 0)\ ,\\
e^{\mu}_{(3)}|_- & = & (0, 0, g, 0)\ ,
\end{eqnarray}
with the prime as always denoting differentiation with respect 
to $\xi _+$. We note that $\cal M^+$ becomes the Schwarzschild space--time 
when $c_n=0$ $(n=0, 1, 2, \dots )$. In this case, by (\ref{3.7}), 
$f_+=1-\xi ^2_+$ and so (\ref{3.5}) becomes $r_+=k\,r\ ,\ \xi _+=\xi\ $
 in agreement with (\ref{2.25}). We note that now $g=k^{-1}$ in 
(\ref{3.5}) and so (3.14)--(3.16) reduce to (2.27)--
(2.29) in this case. Next we take
\begin{equation}\label{3.17}
{}^+N_{\mu}=\left (0, 0, 1, \frac{c_+}{2}\right )\ ,
\end{equation}
with $c_+$ given by (\ref{3.2}) and now (\ref{2.31}) for ${}^-N_{\mu}$ 
is replaced by
\begin{equation}\label{3.18}
{}^-N_{\mu}=\left (-r_+k^2g\,g', 0, g^{-1}, \frac{1}{2\,g}\,
\left (1-\frac{2\,m}{g\,r_+}+(g')^2f_+\right )\right )\ .
\end{equation} 
We can now calculate the components of the transverse extrinsic curvature 
on the plus and minus sides of ${\cal N}$ using the formula (\ref{2.33}). The only 
non--vanishing components ${}^+{\cal K}_{ab}$, in intrinsic coordinates 
$\{\xi ^a\}=\{\xi _+, \phi , r_+\}$ with $a=1, 2, 3$, on ${\cal N}$ are
\begin{equation}\label{3.19}
{}^+{\cal K}_{11}=f_+^{-2}{}^+{\cal K}_{22}=-\frac{m}{f_+}-\frac{r_+f_+''}
{4\,f_+}\ ,
\end{equation}
and the non--vanishing components ${}^-{\cal K}_{ab}$ are
\begin{eqnarray}\label{3.20}
{}^-{\cal K}_{11} & = & -\frac{m}{g^3f_+}-r_+\,\frac{g''}{g}-r_+
\left [\frac{g'}{g}\,\left (\frac{k'}{k}+\frac{f'}{2\,f}+\frac{g'}
{2\,g}\right )-\frac{1}{2\,g^2f_+}\right ]\ ,\\
{}^-{\cal K}_{22} & = & -\frac{m\,f_+}{g^3}-r_+f_+^2\left [\frac{g'}{g}\,\left (\frac{k'}{k}+\frac{f'}{2\,f}+\frac{g'}
{2\,g}\right )-\frac{1}{2\,g^2f_+}\right ]\ ,
\end{eqnarray}
with $f=1-\xi ^2$, and with the prime denoting differentiation with 
respect to $\xi _+$ and $d/d\xi _+=(k\,g)^{-2}d/d\xi$ by (\ref{3.5}) and 
(\ref{3.6}). Next the jump $\sigma _{ab}$ in the transverse extrinsic 
curvature (\ref{2.34}) is calculated, followed by the surface stress--energy 
tensor $S_{\mu\nu}$ given by (\ref{2.36}). This again has the form given 
in (\ref{2.38}) but with the energy density (\ref{2.39}) replaced by
\begin{equation}\label{3.22}
\rho =\frac{\sigma}{16\,\pi}=\frac{1}{4\pi\,r_+^2}\,
\left (\frac{m}{g^3}-m\right )\ .
\end{equation}
To proceed further we need $g(\xi _+)$ given by (\ref{3.6}) and 
this requires $f_+(\xi _+, 0)$. Assuming that the sequence of small 
constants $\{c_n\}$ appearing as coefficients in the expansion (\ref{3.9}) 
is {\it decreasing} we shall take as an {\it approximation} to 
$f_+(\xi _+, 0)$, using (\ref{3.10}),
\begin{equation}\label{3.23}
f_+=1-\xi _+^2+c_0\,(1-\xi _+^2)^2+O_2\ ,
\end{equation}
where we are treating the constant $c_0$ as small of first order 
$\left (O_1\right )$ and the constant $c_1$ as small of second order 
$\left (O_2\right )$ etc.. Now (\ref{3.5}) gives the following 
approximate relation between $\xi _+$ and $\xi$ on $u=0$ (we assume 
that $\xi =\pm 1$ corresponds to $\xi _+=\pm 1$ respectively) 
\begin{equation}\label{3.24}
\xi =\xi _+-c_0\,\xi _+(1-\xi _+^2)+O_2\ .
\end{equation}
It now follows from (\ref{3.6}) that $g(\xi _+)$ is given approximately 
by
\begin{equation}\label{3.25}
g(\xi _+)=\gamma\,(1-v\,\xi _+)\,\left [1+\frac{c_0(1-3\,\xi _+^2)}
{2\,(1-v\,\xi _+)}+\frac{c_0\,v\,\xi _+(1+\xi _+^2)}
{2\,(1-v\,\xi _+)}+O_2\right ]\ .
\end{equation}
This must now be substituted into (\ref{3.22}) and the total 
energy $E_+$ of the light--like shell measured by the distant observer 
who sees the mass $m$ suddenly stop is then found from
\begin{equation}\label{3.26}
E_+=\frac{1}{4\pi}\,\int_{0}^{2\pi}d\phi\,\int_{-1}^{+1}
m\,(g^{-3}-1)\,d\xi _+\ .
\end{equation}
This works out as 
\begin{equation}\label{3.27}
E_+=m\,(\gamma -1)+\frac{3\,m\,c_0}{4\,\gamma ^3v^3}\,
\left [\frac{2\,v\,(5v^2-3)}{3\,(1-v^2)^2}+\log\left (\frac{1+v}{1-v}
\right )\right ]+O_2\ .
\end{equation}
The coefficient of $c_0$ here is always positive for $0<v<1$ and 
vanishes if $v=0$. In the special case of (\ref{3.23}) when
$f_+=1-\xi _+^2$ only the first term in  (\ref{3.27}) survives
and thus  (\ref{2.41}) is recovered.
The first term on the right hand side of (\ref{3.27}) 
is the kinetic energy of the mass $m$ before stopping and it would be 
natural in this model to assume that $c_0<0$ so that some of the 
original kinetic energy is available to be converted into gravitational 
radiation. The coefficient of $\delta (u)$ in the Weyl tensor of 
$\cal M^-\cup\cal M^+$ is non--vanishing and has a radiative part 
\cite {8}, denoted $\hat\Psi _4$ in Newman--Penrose notation, indicating 
the presence of an impulsive gravitational wave in the signal 
with history ${\cal N}$. Here $\hat \Psi _4$ is given by
\begin{equation}\label{3.28}
\hat\Psi _4=\frac{3\,c_0}{r_+}\,(1-\xi _+^2)+O_2\ .
\end{equation}
Thus with $c_0<0$ one can visualise that some of the kinetic energy 
in the original moving mass $m$ is converted into an impulsive 
gravitational wave accompanying the relativistic fireball (light--like shell) 
and the rest supplies the energy for the gravitational radiation present 
in the ring--down phase.

\setcounter{equation}{0}
\section{Discussion}\indent
The line--element induced on the 2--surfaces $r_+=1\ ,\ u={\rm constant}
\geq 0$ in the space--time with line--element (\ref{3.1}) is given 
by
\begin{equation}\label{4.1}
dl^2_+=f_+^{-1}d\xi _+^2+f_+d\phi ^2\ ,
\end{equation}
with $f_+$ given by (\ref{3.10}). It is easy to see that this 2--surface 
can be embedded in three dimensional Euclidean space by rotating a 
curve in the plane $z=0$ the $x$--axis  
given parametrically by
\begin{eqnarray}\label{4.2}
x(\xi _+) & = & \xi _+\,\left\{1-\frac{c_0}{2}\,
(1+\xi _+^2)\,{\rm e}^{-2u/m}+O_2\right \}\ ,\\
y(\xi _+) & = & \sqrt{1-\xi _+^2}\,\left\{1+\frac{c_0}{2}
\,(1-\xi _+^2)\,{\rm e}^{-2u/m}+O_2\right\}\ ,
\end{eqnarray}
with $-1\leq\xi _+\leq +1$. The distance from the origin to any point 
on this 2--surface is
\begin{equation}\label{4.4}
R(\xi _+)=\sqrt{x^2+y^2}=1-c_0\,P_2(\xi _+)\,{\rm e}^{-2u/m}+O_2\ ,
\end{equation}
with $P_2(\xi _+)$ the Legendre polynomial of degree 2 in the 
variable $\xi _+$. The sign of the small parameter $c_0$ was 
determined to be negative on physical grounds following (\ref{3.27}) 
and so we see from (\ref{4.4}) that this means that in the 
ring--down phase $u>0$ the wave fronts of the gravitational waves 
are {\it prolate} spheroids with a common
axis of symmetry coinciding with 
the original direction of motion of the moving mass $m$.

\noindent
\section*{Acknowledgment}\noindent
This collaboration has been funded by the Minist\`ere des Affaires 
\'Etrang\`eres, D.C.R.I. 220/SUR/R.

\end{document}